\newcommand{\be}{\begin{equation}}
\newcommand{\ee}{\end{equation}}
\newcommand{\bee}{\begin{eqnarray}}
\newcommand{\eee}{\end{eqnarray}}
\def\ffrac#1#2{\textstyle{#1\over#2}\displaystyle}
\def\tv{\vartheta_4}
\newcommand{\g}{\gamma_1}
\newcommand{\gb}{\gamma_{11}}
\newcommand{\wed}{(\pi/2)}

\documentstyle[twocolumn,prb,aps]{revtex} 

\input epsf

\begin{document}
\wideabs{ 
\title{EXACTLY SOLVED LATTICE MODELS: FROM POLYMER NETWORKS\\ TO
AN ISING MODEL IN A MAGNETIC FIELD}
\author{M. T. Batchelor\\
Department of Mathematics, School of Mathematical Sciences,\\
The Australian National University, Canberra ACT 0200, Australia}
\maketitle
\vskip 5mm
} 

The study of phase transitions and critical phenomena is one of
the cornerstones of modern theoretical physics. Away from the
critical point, correlations between individual units decay
exponentially with separation. Here the correlation length defines
a characteristic scale. At the critical point, where
the correlation length is infinite and
two or more macroscopic phases become indistinguishable, 
there is long-range power-law decay -- the hallmark of criticality.
Critical points are `scale free', there being no characteristic scale
associated with a simple power law.\cite{Stanley}

Our general understanding of phase transitions and critical phenomena has 
been greatly enhanced by the study of exactly solved lattice models
in statistical mechanics.\cite{Baxter}
Here I give a brief account of some recent developments.
These involve (i) the most general network of
self-avoiding polymer chains attached to a boundary,\cite{BBO} 
and (ii) a model in the same universality class as the two-dimensional
Ising model in a magnetic field.\cite{BSa,BSb}    
The common feature in both studies is an exact solution by means
of the Bethe Ansatz.

\begin{center}
{\bf Polymer Networks}
\end{center}

Progress in polymer physics has been
significantly advanced by the relation with 
critical phenomena.\cite{deGennes}
Long polymer chains in a solvent can be described by
self-avoiding walks (SAWs) on a lattice.
Of particular utility is the correspondence between 
SAWs and the $O(n)$ model in the limit $n \to 0$.
It transpires that this latter model can be solved exactly 
on the honeycomb lattice at criticality.

The investigation of the configurational properties of SAWs in
the vicinity of a boundary already dates back some twenty 
years.\cite{BGMTW78} The SAW is restricted to say the upper half plane
and originates on the boundary. The surface interaction
is represented by an energy, $\varepsilon$,
associated with each contact between the polymer and the surface. The
Boltzmann weight for a configuration of the polymer is given by
$\kappa^m = e^{ m \varepsilon /k_B T}$, where $T$ is the temperature of the
solvent and $m$ is the number of contacts with the surface.  At some
critical temperature, $T_a$, the polymer becomes adsorbed onto the
surface.\cite{HTW,EKB} For $T>T_a$
the polymer is in a desorbed phase where it extends a large distance
into the solvent above the surface. For 
$T<T_a$ the polymer is in an adsorbed phase.
The configurational properties are phase dependent.

The $O(n)$ model has been solved for a number of different boundary 
conditions.\cite{Bfuk} In the terminology of surface critical
phenomena these boundary conditions correspond to the `ordinary',
`special' and `extraordinary' transitions. 
The critical adsorption temperature, $T_a$, for SAWs corresponds 
to the special transition, whilst the ordinary
transition corresponds to SAWs in the presence of an effectively
repulsive surface. The extraordinary transition has also been 
discussed.\cite{BC}

This model for polymer adsorption has been studied on a number of
lattices via different techniques.\cite{DL93,E93}  
The advantage of the present approach is that the relevant quantities
have been calculated exactly on the honeycomb lattice. These
results can then be combined with general arguments\cite{DL93,E93,C96}
from scaling and conformal invariance to yield the
configurational exponents.

\begin{figure}
\centerline{
\epsfxsize=2.5in
\epsfbox{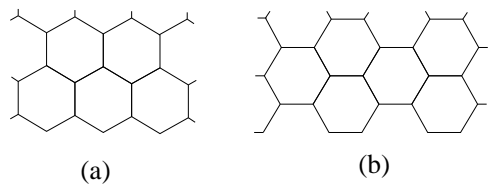}}
\vskip 3mm
\caption{The two orientations of the honeycomb lattice.}
\label{orient}
\end{figure}

The critical adsorption temperature depends on the particular
orientation of the honeycomb lattice. 
For the boundary orientation of Fig.~\ref{orient}(a) the critical
adsorption temperature is known to be given by \cite{BY95a}
\be
\exp \left( \frac{\epsilon}{k T_a} \right) =
{1 + \sqrt 2} = 2.414 \ldots  \label{tempa}
\ee
For the orientation of Fig.~\ref{orient}(b) we find\cite{BBO} 
\be
\exp \left( \frac{\epsilon}{k T_a} \right) =
{{\sqrt{{\frac{2 + {\sqrt{2}}}{1 + {\sqrt{2}} -
{\sqrt{2 + {\sqrt{2}}}}}}}}} = 2.455 \ldots \label{tempb}
\ee
Both results follow from the boundary Boltzmann weights of the
exactly solved $O(n)$ model at $n=0$ and have been confirmed
via series expansion techniques.\cite{BBO,BO96}
However, being specific to the honeycomb lattice, the 
adsorption temperatures are non-universal quantities.
Of most interest are the universal configurational exponents. 

\begin{figure}[h]
\centerline{
\epsfxsize=2.2in
\epsfbox{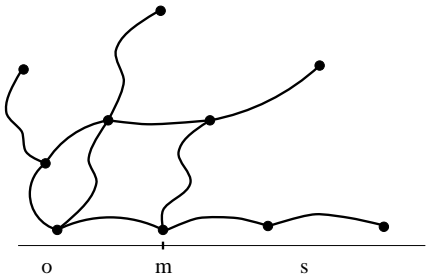}}
\vskip 3mm
\caption{A network of ${\cal N} = 11$ chains attached to the mixed o-s
boundary.}
\label{net}
\end{figure}

Our chief result concerns the
number of configurations ${\cal Z}_G$ of the monodisperse network of
${\cal N}$ self-avoiding polymer chains of length $S$ depicted in 
Fig.~\ref{net}. The network is attached to the boundary. 
The boundary conditions differ on each side of the origin marked m.
To the left is the ordinary boundary condition and to the right
is the special boundary condition.
${\cal Z}_G$ scales as
\be
{\cal Z}_G \sim \mu^{{\cal N} S} S^{\gamma_G -1} \quad \mbox{as}
\quad  S \to \infty,
\ee
where $\gamma_G$ is a universal exponent.
Here $\mu$ is the lattice-dependent connective constant for SAWs.

The general scaling and conformal invariance arguments\cite{DL93,E93,C96} 
can be extended to the mixed boundary. The key ingredients are the
scaling dimensions for each of the bulk, ordinary and special
boundary conditions, all of which have been derived from the
exact Bethe Ansatz solution of the honeycomb lattice model. 
In this way the universal exponent\cite{BBO}
\begin{eqnarray}
\gamma_G = \ffrac{1}{4} &+& \ffrac{1}{64} \sum_L  n_L (2-L) (9L+50)
\nonumber \\
                       &-& \ffrac{1}{32} \sum_L n_L^{\rm o}
       (9L^2 + 22 L - 24) \nonumber \\
  &-&  \ffrac{1}{32} \sum_L n_L^{\rm m} (9L^2 + 10 L - 24)
\\
  &-&   \ffrac{1}{32} \sum_L n_L^{\rm s} (9L^2 -2 L - 16)
   \label{g} \nonumber
\end{eqnarray}
governing the asymptotic number of configurations of
the most general planar network is obtained.

The numbers $n_L$ are topological indices describing any given network.
For example, $n_L$ is the number of $L$-leg vertices in the bulk,
$n_L^{\rm o}$ is the number of $L$-leg vertices attached to the
ordinary surface, $n_L^{\rm m}$ the number at the origin, etc.
In each case there can be $L \ge 1$ vertices. 
The network depicted in Fig.~\ref{net} has  
$n_1 = 3$, $n_3 = 2$, $n_4 = 1$, $n_3^{\rm o} = 1$,
$n_3^{\rm m} = 1$, $n_1^{\rm s} = 1$ and $n_2^{\rm s} = 1$.
For mixed boundaries there is only one $L$-leg vertex emanating
from the origin, thus $n_L^{\rm m} = 1$.

All previously known examples follow from this most general formula.  
The exponents for a pure ordinary surface \cite{DS86} are recovered
with $n_L^{\rm m} = n_L^{\rm s} = 0$. 
For the pure special surface
we take $n_L^{\rm m} = n_L^{\rm o} = 0$.
For a single walk in the half-plane,
$\gamma_1^{\rm o}=\frac{61}{64}$ and $\gamma_1^{\rm s}=\frac{93}{64}$. 
If the walk also terminates on the boundary, 
$\gamma_{11}^{\rm o}=-\frac{3}{16}$ and 
$\gamma_{11}^{\rm s}=\frac{13}{16}$ 
(using the standard notation).
These exponents satisfy Barber's scaling law,
$2\gamma_1 - \gamma_{11} = \gamma + \nu$,\cite{Bar} where
$\gamma = \frac{43}{32}$ and $\nu = \frac{3}{4}$.\cite{DL93,E93,C96}

\begin{figure}[h]
\centerline{
\epsfxsize=2.0in
\epsfbox{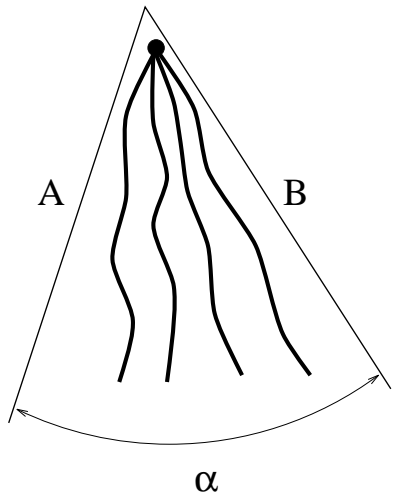}}
\vskip 3mm
\caption{Star polymer in a wedge of angle $\alpha$ with general 
boundaries A and B.}  
\label{star}
\end{figure}

The network can be tied in a wedge of arbitrary angle $\alpha$.
Obtaining the wedge network exponents $\gamma_G(\alpha)$
involves a conformal map of the wedge to the half-plane.
As an example, consider an $L$-leg star polymer
confined to a wedge with o-o, s-s or o-s surfaces (see
Fig.~\ref{star}). The $\alpha$-dependent exponents are given by\cite{BBO}
\bee
\gamma_L^{\rm o}(\alpha) &=& 1 + \frac{27 L}{64} -
\frac{3 \pi L (3L+2)}{32 \alpha} \, , \label{ds} \\
\gamma_L^{\rm s}(\alpha) &=& 1 + \frac{27 L}{64} -
\frac{ 9 \pi L (L-2) + 8 \pi}{32 \alpha} \, , \\
\gamma_L^{\rm m}(\alpha) &=& 1 + \frac{27 L}{64} -
\frac{3 \pi L (3L-2)}{32 \alpha} \, .
\eee
The o-o result (\ref{ds}) is that obtained earlier.\cite{DS86}

As particular examples, consider
a single SAW confined to a quarter of the plane and emanating from 
the $90^\circ$ corner.
In this case the above formulae reduce to
$\gamma_1^{\rm o}(\ffrac{\pi}{2}) = \ffrac{31}{64}$,
$\gamma_1^{\rm s}(\ffrac{\pi}{2}) = \ffrac{95}{64}$,
$\gamma_1^{\rm m}(\ffrac{\pi}{2}) = \ffrac{79}{64}$.
The exponents differ if the walk terminates on either boundary.
In that case we have 
$\gamma_{11}^{\rm o}(\ffrac{\pi}{2}) = -\ffrac{21}{32}$ for the o-o corner
and $\gamma_{11}^{\rm s}(\ffrac{\pi}{2}) = \ffrac{27}{32}$ for the s-s corner.
For the o-s corner the walk can terminate on either the o side,
with $\gamma_{11}^{\rm mo}(\ffrac{\pi}{2}) = \ffrac{3}{32}$,
or on the s side, with 
$\gamma_{11}^{\rm ms}(\ffrac{\pi}{2}) = \ffrac{19}{32}$.

A comprehensive study of a linear chain has been undertaken 
by exact enumeration on the honeycomb lattice, with various 
attachments of the walk's ends to the surface, in wedges of 
angles $\pi/2$ and $\pi$, with general mixed
boundary conditions.\cite{Deb,BO96,BBO}
A comparison between the numerical estimates of the 
$90^\circ$ wedge exponents discussed above and the exact 
values are given below.
The exponent estimates are clearly seen to be in excellent agreement
with the predicted values. The verification of the special exponents
for the $90^\circ$ wedge involved the implicit verification of both the
adsorption temperatures, (\ref{tempa}) and (\ref{tempb}).

\vskip 5mm

\centerline{
\fbox{
$
\begin{array}{rclrcl}
\g^{{\rm o}}\wed &=&0.4843(9)&(\frac{31}{64}&=&0.484375)\\
\gb^{{\rm o}}\wed&=&-0.655(3)&(-\frac{21}{32}&=&-0.65625)\\
\g^{{\rm s}}\wed&=&1.482(8)&(\frac{95}{64}&=&1.484375)\\
\gb^{{\rm s}}\wed&=&0.85(1)&(\frac{27}{32}&=&0.84375)\\
\g^{\rm m}\wed&=&1.233(6)&(\frac{79}{64}&=&1.234375)\\
\gb^{\rm mo}\wed&=& 0.09(1)&(\frac{3}{32}&=&0.09375)\\
\gb^{\rm ms}\wed&=& 0.596(7)&(\frac{19}{32}&=& 0.59375)\\
\end{array}
$
}}

\vskip 5mm

\begin{center}
{\bf Magnetic Ising Model}
\end{center}

The two-dimensional Ising model in a
magnetic field continues to defy an exact solution.
However, a solvable two-dimensional lattice model -- the 
dilute $A_3$ model -- in the same 
universality class as the Ising model in a magnetic field 
has been found.\cite{WNS} 
This model is also known as the $E_8$ lattice realisation of the 
Ising model in a magnetic field $h$ at $T=T_c$. The model
is a three-state interaction-round-a-face model. The Boltzmann
weights, which are too complicated to reproduce here, 
are given in terms of elliptic theta functions. The elliptic nome 
plays the role of a variable magnetic field.
The free energy has been derived,\cite{WNS} the
singular part of which behaves as
\be
f \sim h^{1 + 1/\delta} \quad \mbox{as} \quad h \to 0,
\ee
where the magnetic exponent $\delta=15$.\cite{WNS}
This was the first time that this exponent had been obtained
without the use of scaling relations between critical exponents.\cite{note}
The latest experimental estimate of the Ising magnetic
exponent, on an atomic layer of ferromagnetic iron deposited on a
substrate, is $\delta = 14 \pm 2$.\cite{exp} 
The excess Ising magnetic surface exponent $\delta_s=-\frac{15}{7}$ has 
been derived in a similar way from the excess surface free energy.\cite{BFZ}
This result is also in agreement with that predicted by proposed scaling 
relations between bulk and surface critical exponents.

There is a very intimate relation between exactly solved lattice
models and integrable quantum field theory.
In a remarkable paper, Zamolodchikov \cite{Zb} considered the 
magnetic perturbation of the $c=\frac{1}{2}$ conformal field theory 
and showed that there are a number of nontrivial
local integrals of motion and thus an integrable field theory.
He then conjectured the $S$-matrix and mass spectrum of this field
theory. The masses coincide with the components
of the Perron-Frobenius vector of the Cartan matrix of the Lie algebra
$E_8$. Their ratios are 
\begin{equation}
\begin{array}{ll}
m_2/m_1 = 2 \cos \frac{\pi}{5}  & = 1.618~033\ldots \\
m_3/m_1 = 2 \cos \frac{\pi}{30} & = 1.989~043\ldots \\
m_4/m_1 = 4 \cos \frac{\pi}{5} \cos \frac{7\pi}{30} & = 2.404~867\ldots \\
m_5/m_1 = 4 \cos \frac{\pi}{5} \cos \frac{2\pi}{15} & = 2.956~295\ldots \\
m_6/m_1 = 4 \cos \frac{\pi}{5} \cos \frac{\pi}{30}  & = 3.218~340\ldots \\
m_7/m_1 = 8 \cos^2 \frac{\pi}{5} \cos \frac{7\pi}{30} & = 3.891~156\ldots \\
m_8/m_1 = 8 \cos^2 \frac{\pi}{5} \cos \frac{2\pi}{15} & = 4.783~386\ldots
\end{array}
\label{masses}
\end{equation}

These masses are expected to be present in the scaling limit of the
Ising model in a magnetic field at $T=T_c$. The conjectured $S$-matrix
and the mass spectrum were later confirmed by a thermodynamic
Bethe Ansatz calculation on the Hamiltonian formulation of the
lattice model in the same universality class.\cite{BNW} 

Our approach has been to derive the correlation lengths $\xi_j$ of the
lattice model using an exact perturbative approach based on the
Bethe Ansatz solution.\cite{BSa,BSb}
The input to our calculations are the various root distributions
revealed by numerical investigation.\cite{BNW,GN} 
The calculations are long and tedious, involving perturbation
in an exact manner from the high magnetic field limit.
Our final result is that the elementary masses appearing in the
eigenspectrum are of the form  
\begin{equation}
m_j = \xi_j^{-1} = 2 \sum_a \log \frac{
\tv(\frac{a\pi}{60}+\frac{\pi}{4},h^{8/15})}
{\tv(\frac{a\pi}{60}-\frac{\pi}{4},h^{8/15})},
\label{cor}
\end{equation}
where $\tv$ is a standard elliptic theta function.
Here $h$ is the elliptic nome with $h=0$ at the critical point
and $h=1$ in the high field limit. 
The numbers $a$ as tabulated below have already appeared for the
related Hamiltonian.\cite{MO}
We thus directly see the $E_8$ structure appearing in the
transfer matrix eigenspectrum of the lattice model.\\

\centerline{
\begin{tabular}{|c|l|}
\hline
$j$&$~~a$\\
\hline
1   &        1, 11 \\
2   &        7, 13 \\
3   &        2, 10, 12 \\
4   &        6, 10, 14 \\
5   &        3, 9, 11, 13 \\
6   &        6, 8, 12, 14 \\
7   &        4, 8, 10, 12, 14\\
8   &        5, 7, 9, 11, 13, 15 \\
\hline
\end{tabular}
}
\vskip 5mm

In particular, as $h \to 0$,
\begin{equation}
m_j \sim 8\, h^{8/15}  \sum_a \sin \case{a\pi}{30}.
\label{mo}
\end{equation}
This is the formula obtained previously.\cite{MO}
The $E_8$ masses in (\ref{masses}) are recovered from (\ref{mo}) by
virtue of trig identities. The mass ratios show remarkably little variation
as a function of $h$.
In the high field limit the mass ratios are given exactly by
$1 \case{2}{3}$, 2, $2 \case{1}{2}$, 3, $3 \case{1}{3}$, 4, 5,
which are to be compared with the $h=0$ results (\ref{masses}).

Each of the masses tend to zero at criticality. The related correlation
lengths diverge with magnetic correlation length exponent
$\nu_h = \frac{8}{15}$. This exponent satisfies the  
general scaling relation
$2 \nu_h = 1 + 1/\delta$, which follows from the thermodynamic
relation 
\begin{equation}
f \, \xi^2 \sim \mbox{constant},
\end{equation}
which is expected to hold near criticality.
Here the constant can easily be evaluated. 
The singular part of the free energy scales like
\begin{equation}
f \sim 4 \sqrt 3 \, \frac{\sin{\pi \over 5}}{\cos{\pi \over 30}}
                   \, h^{16/15}
\quad \mbox{as} \quad h \to 0 \, .
\end{equation}
On the other hand, from (\ref{cor}) 
\begin{equation}
\xi_1 \sim {1 \over 8 \sqrt 3 \, \sin{\pi \over 5}} \, h^{-8/15}
\quad \mbox{as} \quad h \to 0 \, .
\end{equation}
Combining these results gives\cite{BSa} 
\begin{equation}
f \, \xi_1^2 = {1 \over 16 \sqrt 3 \sin{\pi \over 5} \cos{\pi \over 30}}
             = 0.061~728~589 \ldots 
\end{equation}
This is the universal magnetic Ising amplitude.
It has been predicted earlier by other means.
Namely by thermodynamic Bethe Ansatz calculations based on the
$E_8$ scattering theory\cite{Zun,F} (see also Ref.~\cite{DM}). 
Here it is obtained for an explicit lattice model.

{~}\\
It is a pleasure to thank Debbie Bennett-Wood, Aleks Owczarek 
and Katherine Seaton for their collaboration in this work.
Financial support from the Australian Research Council is
gratefully acknowledged.

   
\end{document}